\documentclass[aps,prb,twocolumn,superscriptaddress,showpacs]{revtex4-1}

\usepackage{graphicx}
\usepackage{amsmath}
\usepackage{graphics}
\usepackage{exscale}
\usepackage{epsfig}
\usepackage{color}

\begin{document}

\preprint{APS/123-QED}

\title{Kondo effect in Lieb's ferrimagnetic system on the T-shaped bipartite lattice
}

\author{Masashi Tokuda}
\email{tokuda@meso.phys.sci.osaka-u.ac.jp}
\affiliation{Dept. of Physics, Graduate School of Science,Osaka University, Toyonaka, Osaka
560-0043 Japan}
%
\author{Yunori Nishikawa}
\email{nishikaway@osaka-cu.ac.jp, nisikawa@sci.osaka-cu.ac.jp}
%
\affiliation{Dept. of Physics, Graduate School of Science, Osaka City University, Sumiyoshi-ku,
Osaka 558-8585 Japan}
\affiliation{Nambu Yoichiro Institute of Theoretical and Experimental Physics, Osaka City University, Sumiyoshi-ku,
Osaka 558-8585 Japan}


\date{\today}

\begin{abstract}
The minimal ferrimagnetism 
by Lieb's theorem
 emerges on the T-shaped bipartite 
lattice composed of four sites, which can be realized 
experimentally, just as Nagaoka ferromagnetism has been demonstrated experimentally using a quartet quantum-dot
(J.P.Dehollain {\it et al}., Nature {\bf 579}, 528 (2020).).
In this paper, the Kondo effect on this ferrimagnetism is theoretically studied.
The magnetic moment $S=1$ is screened in two steps by the Kondo effect
 and the series conductance $g_{s}$ is strongly suppressed to $g_{s}\simeq 0$, while the parallel conductance $g_{p}$ has the maximum value $g_{p}\simeq 4e^{2}/h$. 
The robustness of these properties against a parameter change
 toward reducing the Lieb's ferrimagnetism is also discussed, 
showing the scenarios for entanglement of the degrees of freedom toward the ground state.

\end{abstract}

\pacs{72.10.F,72.10.A,73.61,11.10.G}

\maketitle


\section{Introduction}

Itinerant magnetism has been one of the intriguing and challenging topics in condensed matter
physics, especially in strongly correlated electron systems.
For example, a reliable general theory predicting magnetic
transition temperature for various magnetic materials has yet to be realized.
Alternately, there are some exact theories predicting the existence of
magnetism in the Hubbard models (see, e.g., a text book on this
topic\cite{tasaki}, and the references therein).
The Lieb's theorem for ferrimagnetism in the half-filled and repulsive
Hubbard model on bipartite lattice is one such theory\cite{lieb1,lieb2} 
 and has been examined in many different material systems; for example,
 honeycomb lattice structures such as graphene
\cite{PhysRevLett.71.4389,PhysRevB.79.241407,PhysRevB.80.045410,PhysRevB.82.045409,ezawa2010generation,jaworowski2013disorder,sharifian2019ground}.
By definition, a bipartite lattice is connected and composed of two
sublattices $A$ and $B$, where any bond connecting sites is in different a sublattice.
The Lieb's theorem states that
the ground state of the half-filled and repulsive Hubbard model on bipartite
lattice has the magnetic moment  $S=|N_{A}-N_{B}|/2$ and is
unique up to the spin degeneracy.
Here $N_{A}$ and $N_{B}$ are the number of lattice points in
sublattice $A$ and $B$, respectively.
According to this theorem, the minimal ( and nontrivial \cite{expnon} ) Lieb's
ferrimagnetism
emerges on the T-shaped bipartite lattice composed of four lattice
points decomposed into $N_{A}=3$ and $N_{B}=1$ and has the magnetic 
moment $S=1$, which we focus on in this paper.
Such a small lattice can be experimentally realized
as a quantum-dot array using recent nanotechnology.
Actually, a controllable quartet quantum-dot plaquette has been
fabricated and the Nagaoka ferromagnetism in the Hubbard model on the plaquette lattice has
been demonstrated experimentally \cite{Natn}.
Similarly, an experimental realization of the minimal Lieb's ferrimagnetism
mentioned above would be possible in the near future. 
Incidentally, the Kondo effect, a screening of a magnetic moment in
an itinerant electron system by the many-body effect, has been investigated
using a quantum-dot array connected to leads as itinerant electron
reservoirs
\cite{goldhaber1998kondo,PhysRevLett.81.5225,potok2007observation,cronenwett1998tunable,jeong2001kondo,iftikhar2015two,keller2015universal,sasaki2000kondo,PhysRevLett.104.036804}.
When the minimal Lieb's ferrimagnetism is realized experimentally,
it would be interesting and challenging to investigate the Kondo effect
on such an intriguing magnetism.
%
In this paper, we  theoretically investigate the Kondo effect on
the minimal Lieb's ferrimagnetism on the T-shaped lattice
connected to reservoirs.
Using the numerical renormalization group (NRG) calculation and the local
Fermi liquid theory, we predict the two-step Kondo screening of the ferrimagnetic
moment, and the strongly suppressed and perfect conductivity through the
T-shaped lattice under the Kondo screening, respectively, for two kinds
of configuration. 
The robustness of these properties against a parameter perturbation
toward reducing the Lieb's ferrimagnetism is also predicted. 

%
This paper is organized as follows.
In Sec. \ref{model}, the model and the formulation we use in this paper are presented.
We show our results in Sec. \ref{result}.
First of all, we investigate the isolated Hubbard model on the T-shaped lattice in Sec. \ref{iso}.
After showing the reliability of our method in Sec. \ref{methocheck}, we present our main results in Sec. \ref{main}.
The robustness of our findings against parameter perturbations is discussed in Sec. \ref{robust}.
Section \ref{last} is devoted to the Conclusion.

\section{Model and Formulation}\label{model}
\begin{figure}[h]
\includegraphics[width=0.5\textwidth]{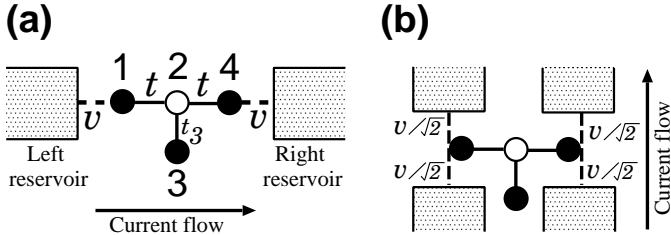}
\caption{Schematic picture of (a) series and (b) parallel configurations}\label{Tzu}
\end{figure}

The model we consider is a Hubbard model on
the T-shaped bipartite lattice decomposed into the sublattice
$A=\{1,3,4\}$ and $B=\{2\}$, 
 which connects two reservoirs at the left (L) and right (R) by the symmetrical tunneling
matrix elements $v$, as illustrated in Fig. \ref{Tzu} (a). 
The Hamiltonian ${\cal H}$ is given by  
$\displaystyle {\cal H}={\cal
H}_{\rm T}+{\cal H}_{\rm res}+{\cal H}_{\rm hyb}$  with
\begin{eqnarray}
{\cal H}_{\rm T}&=&\sum_{\sigma,i\in A,j\in
 B}t_{ij}d^{\dagger}_{i\sigma}d_{j\sigma}+\sum_{i,\sigma}\left(\varepsilon_{d,i}n_{i\sigma}+U_{i}n_{i\uparrow}n_{i\downarrow}\right),\\ 
{\cal H}_{\rm res}&=&\sum_{\nu,k,\sigma}\varepsilon_{\nu
 k}c^{\dagger}_{\nu k\sigma}c_{\nu k\sigma},\\
{\cal H}_{\rm hyb}&=&v\left(d^{\dagger}_{1\sigma}\psi_{L\sigma}+h.c.\right)+v\left(d^{\dagger}_{4\sigma}\psi_{R\sigma}+h.c.\right),
\end{eqnarray}
where $d_{i\sigma}$ annihilates an electron with spin $\sigma$ at the
site-$i$ in the T-shaped lattice, 
characterized by the intersite hopping matrix elements $t_{ij}$ between
site $i\in A$ and $j\in B$, the onsite
energy $\varepsilon_{d,i}$ and the intrasite repulsion $U_{i}$.
Here $n_{i\sigma}\equiv d^{\dagger}_{i\sigma}d_{i\sigma}$ is the number
operator of the electron with spin $\sigma$ at the site-$i$.
The necessary condition for the emergence of the Lieb's ferrimagnetic state is 
a half-filled repulsive Hubbard model on a bipartite lattice, so that
the symmetry of the lattice is not a main factor for emerging the Lieb's ferrimagnetic state.
Therefore, for simplicity 
we assume $\varepsilon_{d}\equiv
\varepsilon_{d,1}=\varepsilon_{d,2}=\varepsilon_{d,4}$, \ $\varepsilon_{3}\equiv\varepsilon_{d,3}$
$t\equiv t_{12}=t_{24}$, $t_{3}\equiv t_{23}$ and $U\equiv
U_{1}=U_{2}=U_{4}$ throughout this paper and set mainly $\varepsilon_{d}=\varepsilon_{3}, \ 
t=t_{3}$ and $U=U_{3}$ unless otherwise stated.
In the reservoir at $\nu (=R,\ L)$, $c^{\dagger}_{\nu k\sigma}$ creates an electron
with energy $\varepsilon_{\nu k}$ corresponding to an one-particle state
$\phi_{\nu k}(r)$ and 
$\psi_{\nu\sigma}=\sum_{k}\phi_{\nu k}(r_{\nu})c_{\nu k\sigma}$
is the field operator of the conduction electron in the reservoir at $r_{\nu}$ where the conduction
electrons in the reservoir mix with the electrons in the site labeled by
$i=1$ (for $\nu=L$) or $i=4$ (for $\nu=R$).
We assume that the hybridization strength $\Gamma\equiv\pi
v^{2}\sum_{k}|\phi_{\nu k}(r_{\nu})|^{2}\delta(\omega-\varepsilon_{\nu k})$
is a constant independent of the frequency $\omega$ and $\nu$, and take
the Fermi energy $\mu$ to be $\mu=0$. Hence, assuming that the
conduction electron in the reservoir has a flat band structure with half
bandwidth $D$, we have $\Gamma=\pi v^{2}/(2D)$.
Our system has inversion symmetry, so that the even and odd parities are
good quantum numbers.
Therefore, it is convenient to introduce the even-parity orbitals
$a_{1\sigma}, \ a_{2\sigma}, \ a_{3\sigma}$
and the odd-parity orbital $b_{1\sigma}$ as follows;
\begin{eqnarray}
a_{1\sigma}&=&\frac{d_{1\sigma}+d_{4\sigma}}{\sqrt{2}},
a_{2\sigma}=d_{2\sigma},
a_{3\sigma}=d_{3\sigma},\\
b_{1\sigma}&=&\frac{d_{1\sigma}-d_{4\sigma}}{\sqrt{2}}.
\end{eqnarray}
The retarded Green's functions for $a_{1\sigma}$ and $b_{1\sigma}$
play an important role for calculating the conductance through the T-shaped
lattice and the averaged electron number in the lattice because the
orbitals $d_{1\sigma}$ and $d_{4\sigma}$ connect to the reservoirs.
Due to the inversion symmetry, at the zero temperature and Fermi energy,
each of these two retarded Green's functions is determined by a single
real parameter, $\kappa_{e}$ or $\kappa_{o}$.
The parameter $\kappa_{p} \ (p=e, o)$ is defined by,
\begin{equation}
\kappa_{p}
=
\frac{{\rm det}K_{p}}{\Gamma {\rm det}K_{p,11}}.\label{kappa-ham}
\end{equation}
Here, 
$K_{p}\equiv -(h^{(0)}_{p}+{\rm Re}\Sigma^{+}_{p}(0))$,
where $h^{(0)}_{p}$ is the matrix composed of the hopping integrals among the
$p$-parity orbitals,
$\Sigma_{p}^{+}(\omega)$ is the self-energy with the $p$-parity and 
$K_{p,11}$ is the matrix obtained by deleting the first row and column
corresponding to the orbital $a_{1\sigma}$ or $b_{1\sigma}$
from the matrix $K_{p}$.
These real parameters determine the phase shifts $\delta_{e}$ and
$\delta_{o}$ corresponding to the angles of these two Green's functions in
the complex plane as follows; $\delta_{e}=\arctan(-1/\kappa_{e})$, $\delta_{o}=\arctan(-1/\kappa_{o})$.
These two phase shifts $\delta_{e}$ and $\delta_{o}$ of the quasi-particles
with even- and odd-parity characterize a local Fermi-liquid
behavior of the whole system described by ${\cal H}$.
The conductance $g_{s}$ in the two-terminal series configuration
illustrated in Fig.\ref{Tzu}(a) and the averaged electron number $n\equiv\langle G|\sum_{i,\sigma}n_{i\sigma}|G\rangle$
in all sites of the ground state $|G\rangle$ are represented
\cite{onh,no} as follows;
\begin{eqnarray}
g_{s}&=&\frac{2e^{2}}{h}\sin^{2}(\delta_{e}-\delta_{o})
, \\
n&=&\frac{2}{\pi}\left(\delta_{e}+\delta_{o}\right).
%
\end{eqnarray}

From the same phase shifts, we can calculate the conductance $g_{p}$ in the
four-terminal
parallel configuration illustrated in Fig.\ref{Tzu} (b) as follows;
\begin{equation}
g_{p}=\frac{2e^{2}}{h}\left(\sin^{2}\delta_{e}+\sin^{2}\delta_{o}\right).
\end{equation}


We perform NRG calculation to determine $\delta_{e}$ and $\delta_{o}$.
In the NRG approach, a sequence of the Hamiltonian $H_{N}$ is
introduced, by carrying out the logarithmic discretization with the
control parameter $\Lambda$ for the continuous conduction bands of the electron reservoirs, and trasforming the discretized electron
reservoirs  as,
\begin{eqnarray}
H_{N}&=&\Lambda^{(N-1)/2}\left(
{\mathcal H}_{\rm T}
+
{\mathcal H}_{\rm NRG:hyb}
+
{\mathcal H}_{\rm NRG:res}^{(N)}
\right),\\
H_{\rm NRG:hyb}&=&\overline{v}\sum_{\sigma}(d_{1\sigma}f^{\dagger}_{0,L\sigma}+d_{4\sigma}f^{\dagger}_{0,R\sigma}+h.c.),\\
H^{(N)}_{\rm
 NRG:res}&=&D\frac{1+1/\Lambda}{2}\sum_{\nu=R,L}\sum_{\sigma}\sum_{n=0}^{N-1}\xi_{n}\Lambda^{-n/2}\nonumber\\
& &\mbox{}\times(f_{n,\nu\sigma}f^{\dagger}_{n+1,\nu\sigma}+h.c.),
\end{eqnarray}
where $f_{n,\nu\sigma}$ annihilates an electron with spin $\sigma$ at
site $n$ in the $\nu$-discretized electron reservoir,  $\overline{v}=\sqrt{2D\Gamma A_{\Lambda}/\pi},
A_{\Lambda}=\frac{1}{2}(1+1/\Lambda)/(1-1/\Lambda)\log\Lambda$, and
\begin{equation}
\xi_{n}
=
\frac
{1-1/\Lambda^{n+1}}
{\sqrt{1-1/\Lambda^{2n+1}}\sqrt{1-1/\Lambda^{2n+3}}}.
\end{equation}

We keep the lowest 3600 eigen states during the NRG iteration process and
set $\Lambda=6$ in our NRG calculations.

We can deduce $\delta_{e}$ and $\delta_{o}$ via $\kappa_{e}$ and
$\kappa_{o}$ from the fixed-point eigen
energies of the 
NRG
calculation.\cite{onh,no}
as follows;
\begin{equation}
\kappa_{p}
=\frac{\overline{v}^{2}}{\Gamma D}
\lim_{N\to\infty}
D\Lambda^{\frac{N-1}{2}}
g_{N}(\epsilon^{\ast}_{p}).
\end{equation}
Here, 
$\epsilon_{p}^{\ast}$ is the quasiparticle energy with the $p$-parity
obtained from the NRG fixed-point eigen energies
, 
 and 
$g_{N}$ is the Green's function for one of the isolated discretized
electron reservoirs.

We have confirmed that the numerical results for the fixed-point
eigenvalues can be mapped onto the energy spectrum of the free
quasiparticles in all parameter sets we have examined, which justifies
the assumption of the local Fermi liquid we have made in our formulation.

Using the NRG flow, we can calculate the impurity (T-shaped lattice) entropy as a function of the discretized temperature 
$T_{N_{\rm nrg}}\equiv \tau\Lambda^{-(N_{\rm nrg}-1)/2}D$ corresponding to
the number $N_{\rm nrg}$ of NRG iterations \cite{kww}. (Here $\tau=O(1)$ is a fitting
constant 
.)

\section{Results}\label{result}
\subsection{Results for the isolated Hubbard model on the T-shaped lattice}\label{iso}
\begin{figure}[h]
\includegraphics[width=0.5\textwidth]{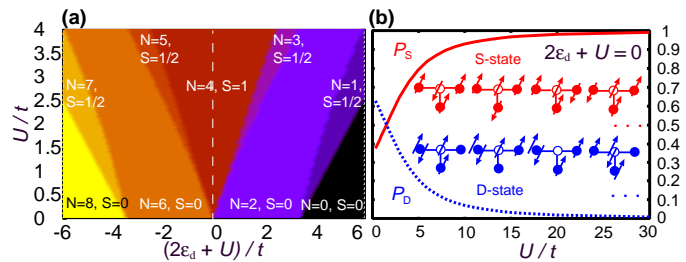}
\caption{ (Color online) (a) The spin quantum number and the electron
 occupation number of the ground state of the 
 isolated Hubbard model on the T-shaped lattice as functions of
 $(2\varepsilon_{d}+U)/t$ and $U/t$. (b) The probabilities $P_{S}$ and
 $P_{D}$ plotted as functions of $U/t$. Examples of S- and D-state are
 presented in this figure.}\label{dia}
\end{figure}

Firstly, we investigate the isolated Hubbard model on the T-shaped lattice before
connecting the electron reservoirs, clarifying the spin state $S$ and
the electron occupation number $N$
 in the model parameter space.
In Fig.\ref{dia}(a), we show the phase diagram of $S$ and $N$
on the model
parameter plane spanned by  $(2\varepsilon_{d}+U)/t$ and $U/t$.
At the half-filled state $N=4$, we confirm that the ground state with
$S=1$ is realized for any positive value of $U$ 
 in our system\cite{PhysRevB.100.224421}, as is predicted by Lieb's theorem.
The region of $N=4$ and $S=1$ becomes wider as $U$ increases.
From the phase diagram of $N$, we easily realize that the phase diagram of $N-4$, the electron occupation number from the
half-filled state, is antisymmetric with respect to the line
$2\varepsilon_{d}+U=0$ (the white dashed line) in the phase diagram.
This is because our system has the electron-hole symmetry.
As a result, the phase diagram of $S$ is symmetric with respect to the line.
Along the line $U=0$ in the phase diagram, the value of $N$ changes by
increments of two, while $S=0$ because two electrons with up and down spin occupy the energy
level crossing the Fermi energy at the same time.

Next, we examine the minimal Lieb's ferrimagnetic state in more detail. 
The basis vectors that span the $N=4$ and $S=1$ states including the minimal Lieb's ferrimagnetic state 
 can be
classified into two types, namely,  S-state and D-state.
A basis vector that consists of only single occupied sites belongs to
the S-state and a basis vector that has a doubly occupied site belongs
to the D-state.
Some examples of the S- and D-state are presented inside Fig.\ref{dia}(b).
A so-called ferrimagnetic state where the spins on each sublattice are in
ferromagnetic order and two spins on the different sublattices are
anti-parallel, is a superposition state of the basis vectors belonging
to the S-state.   
The ground state $|G_{\rm iso}\rangle$ of the isolated Hubbard model
with $N=4$,  $S=1$ (and $S_{z}=1$) is a superposition
state of the basis vectors belonging to the S- and D-state.
To estimate how close the ground state is to the so-called ferrimagnetic state, 
we calculate two probabilities $P_{S}\equiv \langle
G_{\rm iso}|\hat{P}_{S}|G_{\rm iso}\rangle$
and 
$P_{D}\equiv \langle
G_{\rm iso}|\hat{P}_{D}|G_{\rm iso}\rangle$, where $\hat{P}_{S} \ (\hat{P}_{D})$ is the
projection operator to the subspace spanned by the basis vectors
belonging to the S-state (D-state).
The results are shown in Fig.\ref{dia}(b).
The value of $P_{D}$ decreases as $U$ increases because the doubly
occupied state is unfavorable due to the intrasite repulsion $U$.
As a result, $P_{S}$ increases as $U$ increases because of the
constraint condition $P_{S}+P_{D}=1$.

\begin{figure}[h]
\includegraphics[width=0.5\textwidth]{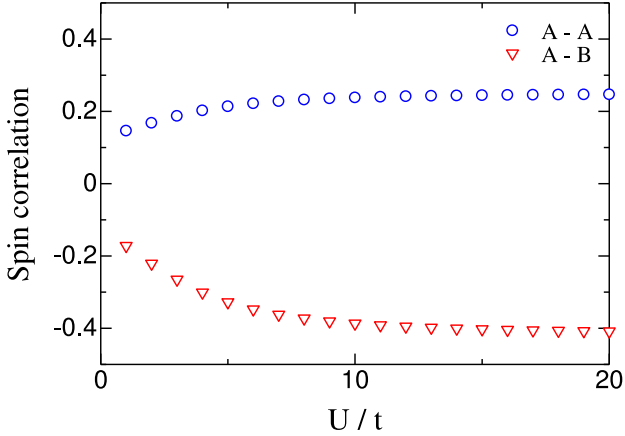}
\caption{ (Color online) 
The spin correlations between two sites in sublattice $A$ (blue circle) and between site-$i$ in sublattice $A$ and
site-$j(=2)$ in sublattice $B$ (red inverted triangle) plotted as functions of $U/t$.
}\label{sc}
\end{figure}

Lastly, we investigate how the spins align in the ground state $|G_{\rm
iso}\rangle$ by
calculating the spin correlations $\langle G_{\rm iso}|{\boldsymbol
S}_{i}\cdot{\boldsymbol S}_{j}|G_{\rm
iso}\rangle$ between site-$i$ and site-$j$ as functions of $U$.
The results are shown in Fig.\ref{sc}, where the blue circles and the
red inverted triangles  represent the results for the spin correlation
between two sites in sublattice $A$ ( the intra-sublattice spin
correlation ), 
and the spin correlations between site-$i$ in sublattice $A$ and
site-$j(=2)$ in sublattice $B$ ( the inter-sublattice spin correlation ), respectively.
These results show that the spins in sublattice $A$ ferromagnetically
align and two spins on the different sublattices are in
antiferromagnetic order.
As $U$ is increased, the values of the intra- and inter- sublattice spin
correlations saturate
to $\frac{1}{4}$ and $-\frac{5}{12}=-0.4166..$, respectively.
These saturation values can be explained as follows.
In the limit of large $U$, the ground state $|G_{\rm
iso}^{U=\infty}\rangle$ is a superposition
state of the basis vectors belonging to the S-state only, as shown in
Fig.\ref{dia}(b) and is expressed by

\begin{eqnarray}
|G_{\rm iso}^{U=\infty}\rangle&=&
\sqrt{\frac{9}{12}}
d_{1\uparrow}^{\dagger}
d_{3\uparrow}^{\dagger}
d_{4\uparrow}^{\dagger}
d_{2\downarrow}^{\dagger}
|0\rangle
-
\sqrt{\frac{1}{12}}
(
d_{1\downarrow}^{\dagger}
d_{3\uparrow}^{\dagger}
d_{4\uparrow}^{\dagger}
d_{2\uparrow}^{\dagger}
|0\rangle \nonumber\\
& &\mbox{ }
+
d_{1\uparrow}^{\dagger}
d_{3\downarrow}^{\dagger}
d_{4\uparrow}^{\dagger}
d_{2\uparrow}^{\dagger}
|0\rangle 
+
d_{1\uparrow}^{\dagger}
d_{3\uparrow}^{\dagger}
d_{4\downarrow}^{\dagger}
d_{2\uparrow}^{\dagger}
|0\rangle 
)\label{Giso}
\end{eqnarray}
, where $|0\rangle$ is the vacuum state.
Using this expression, we obtain the two saturation values, 
$\langle G_{\rm iso}^{U=\infty}|{\boldsymbol
S}_{1}\cdot{\boldsymbol S}_{3}|G_{\rm
iso}^{U=\infty}\rangle=\frac{1}{4}$
and
$\langle G_{\rm iso}^{U=\infty}|{\boldsymbol
S}_{1}\cdot{\boldsymbol S}_{2}|G_{\rm
iso}^{U=\infty}\rangle=-\frac{5}{12}$


From these calculations, it is found that a sizable value of $U(\gg t)$ is required to regard the ground
state as a so-called ferrimagnetic state (we mainly set $U/t=4$). 

%

%
\subsection{Results for noninteracting system}\label{methocheck}
\begin{figure}[h]
\includegraphics[width=0.5\textwidth]{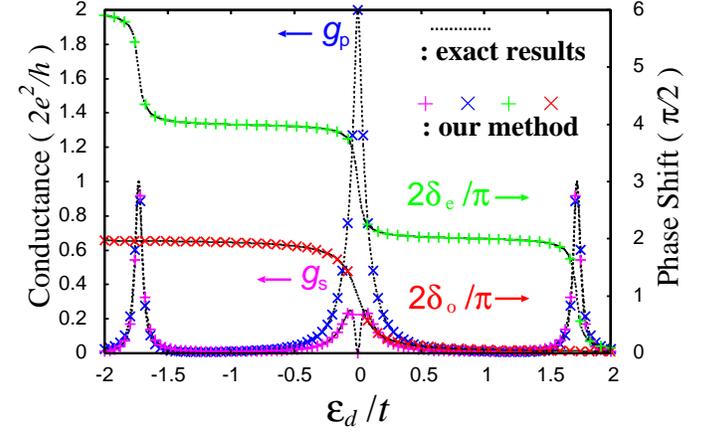}
\caption{ (Color online) Two kinds of conductance $g_{s}, g_{p}$ and the phase shifts $\delta_{e}, \delta_{o}$ obtained using our method are compared with the exact results for $U=0$. Here $\Gamma/t=0.1$. For NRG, we use $\Lambda =6$ and $t/D=0.1$.}\label{u=0}
\end{figure}

Henceforward, we investigate the Hubbard model on the T-shaped lattice connected to the
electron reservoirs.
For $U=0$, changing the value of $\varepsilon_{d}$, we calculate $g_{s}, \ g_{p}, \ \delta_{e}$ and $\delta_{o}$
by using our method and compare
the calculated results with the exact results in 
Fig.\ref{u=0}.
The exact expressions of $\kappa_{e}$ and $\kappa_{o}$ for $U=0$
are easily obtained from Eq.(\ref{kappa-ham}) because
$\Sigma^{+}_{e}(0)=\Sigma^{+}_{o}(0)=0$ and
\begin{equation}
K_{e}=\left(
\begin{array}{ccc}
\varepsilon_{d} & \sqrt{2}t & 0\\
\sqrt{2}t & \varepsilon_{d} & t\\
0 & t & \varepsilon_{d} 
\end{array}
\right),\ \ 
K_{o}=\left(
\begin{array}{c}
\varepsilon_{d} \\
\end{array}
\right).
\end{equation}

Then we obtain the exact expressions of $\delta_{e}$ and $\delta_{o}$
for $U=0$ as follows;
\begin{eqnarray}
\delta_{e}&=&\arctan(-\Gamma(\varepsilon_{d}^{2}-t^{2})/(\varepsilon_{d}(\varepsilon_{d}^{2}-3t^{2}))),\\
\delta_{o}&=&\arctan(-\Gamma/\varepsilon_{d}).
\end{eqnarray}
It is found that the agreement between the results by our method and the
exact results are remarkably consistent even for a rather large value of
the discretization parameter $\Lambda =6$.
 Therefore, the effect of the NRG
discretization on these quantities is negligible and our method for
calculating these quantities is reliable.

Two conductances $g_{s}$ and $g_{p}$ have large values when the electron
occupation number $n=2(\delta_{e}+\delta_{o})/\pi$ changes by two because a pair of electrons with up
and down spins from the reservoir occupy the energy level of the impurity resonating with the Fermi level. 
%

\subsection{Kondo effect}\label{main}
\begin{figure}[h]
\includegraphics[width=0.5\textwidth]{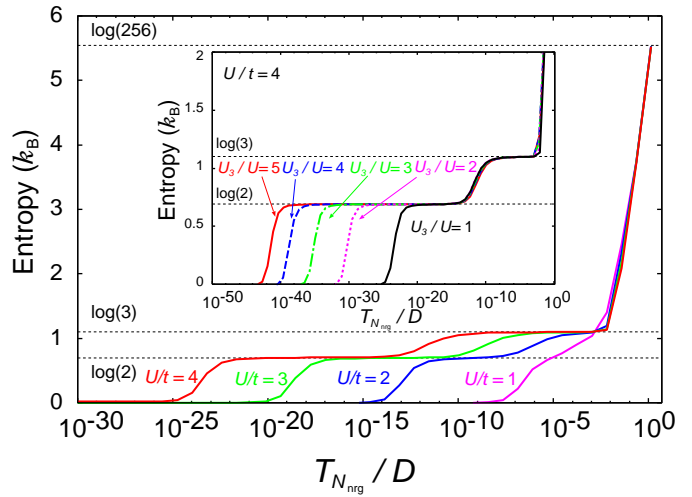}
\caption{ (Color online) Temperature dependencies of the impurity
 entropy for several values of $U$ calculated using NRG energy
 spectrum. (Inset) The $U_{3}/U$ dependencies of the impurity entropy.}\label{entro}
\end{figure}

\begin{figure}[h]
\includegraphics[width=0.4\textwidth]{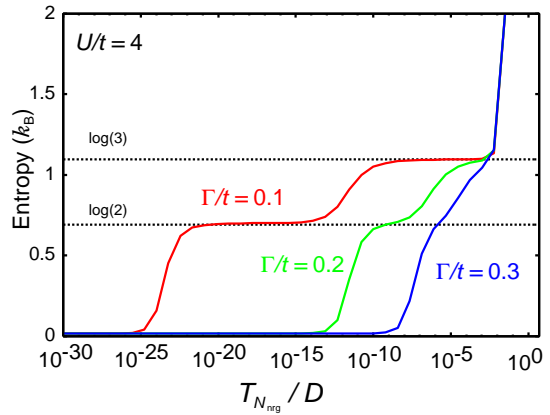}
\caption{ (Color online)$\Gamma$-dependence of the impurity entropy for $U/t=4$}\label{Gentro}
\end{figure}

We next investigate how the Lieb's ferrimagnetic moment $S=1$ 
at the half-filled state ($\varepsilon_{d}/t=-U/(2t)$) is screened by the Kondo effect.
We show the impurity entropy as a function of the temperature, varying
the values of $U$ in Fig.\ref{entro}.
In the high temperature region, the values of the impurity entropy for
all the values of $U$ shown in the figure
are $\log(256)$ because all degrees of freedom $256=2^8$ of the four sites appear
in this region.
Decreasing the value of $T_{N_{\rm nrg}}$ from the high temperature region, we observe
the $\log(3)$-plateau due to the Lieb's ferrimagnetism $S=1$ on the
T-shaped lattice.
This degree of freedom is screened by the Kondo effect in the low
temperature region via the $\log(2)$-plateau and
the Kondo temperature, which is the screening temperature required to reach the singlet
states,
 decreases as the value of $U$ increases.
Therefore, the magnetic moment in the Lieb's ferrimagnetism on the
T-shaped lattice is not screened in one step by the conduction electrons
from the two symmetrically connected reservoirs but is completely screened in two steps with different
energy scales.
In the first Kondo screening ($\log(3) \to \log(2)$), the reduced degrees of freedom $f_{1}$ is $f_{1}\simeq 1(=3-2)$.
Therefore, the partial magnetic moment of the Lieb's ferrimagnetic state
with $S=1$ screened in the first Kondo screening can not correspond to any
positive integer or half-integer magnetic moment $s_{1}$ because
$2s_{1}+1=f_{1}\Longleftrightarrow s_{1}=0$.
To give some insights into the screening mechanism,
 we consider the distribution of the $S=1$ moment on the T-shaped lattice.
The two-step Kondo screening becomes more significant for large $U$ and
small $\Gamma$ 
as shown in Fig.\ref{entro} and Fig.\ref{Gentro}.
Therefore, it is reasonable 
to consider the momentum distribution in large $U$ and small $\Gamma$ limit.
In the limit, the ground state of the isolated Hubbard model on the T-shaped
lattice with $S=1, \ S_{z}=1$ is given by Eq.(\ref{Giso}). 
Using these assumptions, we can calculate the momentum distribution $m_{i}\equiv \langle
G^{U=\infty}_{\rm
iso}|\frac{1}{2}(d_{i\uparrow}^{\dagger}d_{i\uparrow}-d^{\dagger}_{i\downarrow}d_{i\downarrow})|G^{U=\infty}_{\rm
iso}\rangle$ for each site-$i$ of the T-shaped lattice as follows;
$m_{1}=m_{3}=m_{4}=\frac{5}{12}$ and $m_{2}=-\frac{3}{12}$.
The results show that the momentum is mainly distributed among sites-1,3,and 4.
When the reservoirs connect to the T-shaped lattice, 
site-1 and site-4 directly couple to the reservoirs.
Therefore, the partial moments on these two sites can be screened 
directly by the conduction electrons at the first step.
In contrast, site-3 does not connect to the reservoirs directly and thus the partial moment on site-3  has to be screened indirectly by the conduction electron through site-2, which corresponds to the second step of the Kondo screening.
To confirm the discussion mentioned above,
 we increase the value of $U_{3}$, which is the intrasite repulsion
of the site-3 (the most internal site from the reservoirs), 
from $U_{3}=U$, keeping the relation $2\varepsilon_{3}+U_{3}=0$ for the
half-filled state.
The results are shown in the inset of Fig.\ref{entro}.
The shape of the curve of the impurity entropy from the high temperature
region to the beginning of the $\log(2)$-plateau via the
$\log(3)$-plateau is almost insensitive to $U_{3}/U$, while the length
of the $\log(2)$-plateau becomes longer as the value of $U_{3}/U$
increases.
Then the Kondo temperature decreases as the value of $U_{3}/U$
increases.
From these facts, we can confirm that the second screening process corresponds to the screening
of 
the partial magnetic moment distributed on the most internal site-3.

We consider the dependence of the two-step Kondo screening on the
hybridization $\Gamma$, for $U/t=4$.
The impurity entropy for the intermediate coupling $\Gamma=0.2$
retains the structure observed in the weak coupling $\Gamma=0.1$.
For the large hybridization $\Gamma=0.3$, the charge transfer between the T-shaped lattice and the reservoir brings the system into a mixed-valence regime and the typical structures are smeared out.
The Kondo temperature is sensitive to the value of $\Gamma$ and it increases with $\Gamma$.
This is because the large hybridization makes the resonance peaks broad and it reduces effectively the correlation effects.

\begin{figure}[h]
\includegraphics[width=0.5\textwidth]{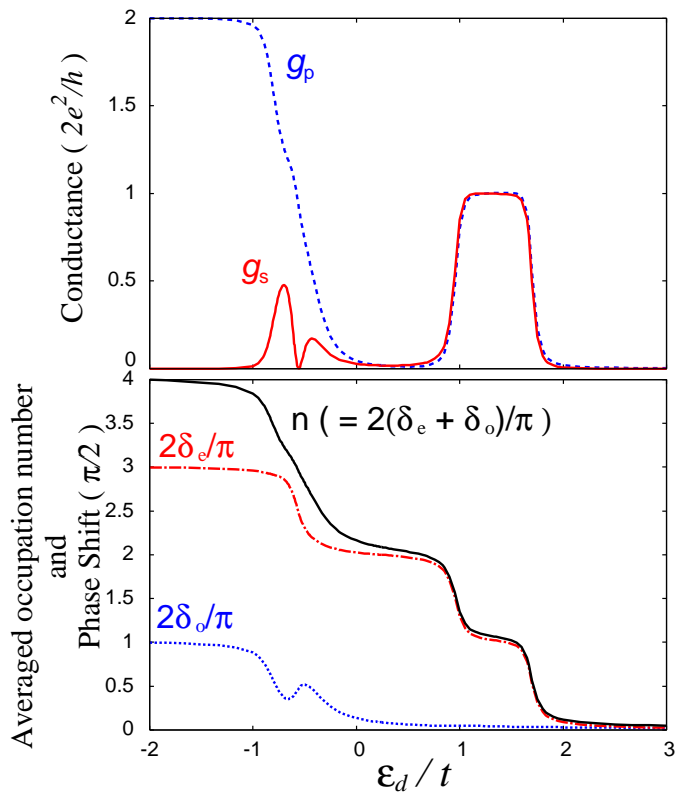}
\caption{ (Color online) Two kinds of conductance $g_{s}, g_{p}$, the
 electron occupation number $n$ in the T-shaped lattice, and the
 phase shifts $\delta_{e}$ and $\delta_{o}$ plotted as functions of $\varepsilon_{d}/t$ for $U/t=4$.}\label{coned}
\end{figure}

To investigate behaviors of two conductances $g_{s}$ and $g_{p}$ under
the Kondo screening of the
 Lieb's ferrimagnetism emerging at the half-filled state shown above, 
we calculate $g_{s}, \ g_{p}$ and $\delta_{e}, \ \delta_{o}$ as
functions of $\varepsilon_{d}$ and show the results in Fig.\ref{coned}. 
Around the half-filled state $\varepsilon_{d}/t=-2(=-U/(2t))$,  
the value of $g_{s}$ is strongly suppressed $g_{s}\simeq 0$ in spite of the existence
of the Kondo screening, while the value of $g_{p}$
reaches its maximum value $g_{p}\simeq 4e^{2}/h$ because $\delta_{e}$ and $\delta_{o}$
respectively have $3\pi/2$- and $\pi/2$-
plateaus around the half-filled state.
One possible reason for this is as follows.
There is a possibility of a residual anti-ferromagnetic correlation
between the site-2 and 3 resulting from the Lieb's
ferrimagnetic state,
 which prevents conductivity in the series
configuration, but causes perfect conductivity in the parallel
configuration.
This is  because the anti-ferromagnetic coupling between site-2
and 3 blocks the branch path for conductivity in the parallel
configuration. 
Increasing the value of $\varepsilon_{d}$ from the half-filled state,
we can see the region where the value of $n$ transitions from 4 to 2 via 3.
In this region, the graph of $g_{s}$ has a double-peak structure because
of the dip structure in the behavior of $\delta_{o}$.
This interesting behavior will be investigated elsewhere because, in
this paper, we focus on the Kondo effect on the minimal Lieb's ferrimagnetism emerging
at the half-filled state. 
Around the region where $n\simeq 1$, 
we can see the typical Kondo plateau of both conductances $g_{s}$ and
$g_{p}$, which is principally caused by the even-parity states. 
%
\subsection{Robustness against parameter perturbation toward reducing
  the ferrimagnetism}\label{robust}
 
\begin{figure}[h]
\includegraphics[width=0.5\textwidth]{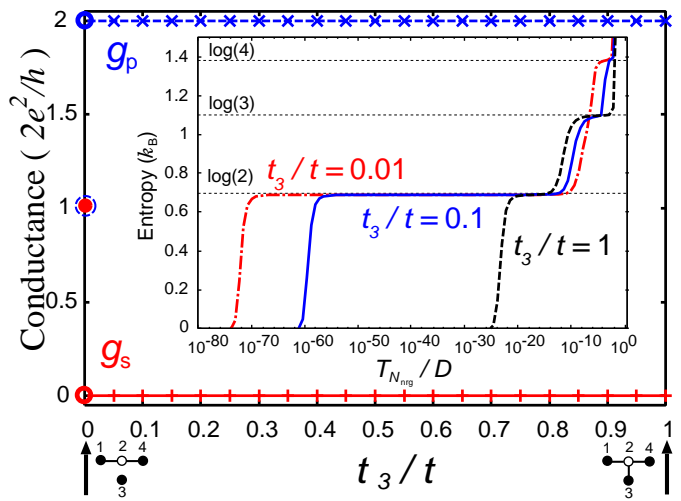}
\caption{ (Color online) Two kinds of conductance $g_{s}$ and  $g_{p}$
 plotted as
 functions of $t_{3}/t$ with discontinuities at $t_{3}/t=0$. (Inset)
 Temperature dependencies of the impurity entropy for $t_{3}/t=0.01$ and $0.1$ compared with the result for $t_{3}/t=1$. }\label{tdep}
\end{figure}

Finally, we study in more detail the behaviors of both conductance $g_{s}$ and $g_{p}$ at
the half-filled state
mentioned above, by setting $t\neq t_{3}$ and reducing the value of $t_{3}/t$. 
At $t_{3}/t=0$, our system is decoupled into the half-filled 3-site Hubbard chain
connected to two electron reservoirs and the isolated site-3 occupied by one
electron.
In this case, the spin $S=1/2$ on the half-filled 3-site Hubbard chain
is screened by the Kondo effect in the ground state, which gives conductivities in both
configurations ($g_{s}\simeq 2e^{2}/h, \ g_{p}\simeq 2e^{2}/h$)\cite{onh} and we have the residual impurity
entropy $\log(2)$ by the degrees of the freedom of the spin $S=1/2$ on
the isolated site-3.
The question arises as to the $t_{3}/t$-dependence of two conductances. 
To answer this question, we calculate $g_{s}$ and $g_{p}$ as functions of $t_{3}/t$ and show the
results in Fig.\ref{tdep}.
We find that the values of $g_{s}$ and $g_{p}$ for $t_{3}/t >0$ keep the
constant values at $t_{3}/t=1$ and change discontinuously only at $t_{3}/t=0$.
Therefore, the behaviors of $g_{s}$ and $g_{p}$ at the half-filled state
are robust against the parameter perturbation toward reducing the Lieb's ferrimagnetism.
We investigate this robustness from $t_{3}/t$-dependence of the impurity
entropy as a function of the temperature.
For $t_{3}/t=0.1$ and $0.01$, the impurity entropies are plotted as functions
of the temperature and are compared with the result for $t_{3}/t=1$ in
the inset of Fig.\ref{tdep}.
The value of $U$ is fixed at $U/t=4$.
We find that the Kondo temperature decreases as the value of $t_{3}/t$ decreases.
For $0.1<t_{3}/t<1$, we confirmed that as the temperature decreased, the
value of the impurity entropy decreases
from the value $\log(256)$ and directly reaches the $\log(3)$-plateau
corresponding to the Lieb's ferrimagnetic state (an entanglement state
of four sites), and converges to zero
via the $\log(2)$-plateau by forming the Kondo singlet state (an
entanglement state of four sites and two reservoirs).
Therefore, the scenario for the entanglement of the degrees of freedom 
($\log(3)\rightarrow \log(2)\rightarrow 0$)
from
the high temperature region toward the ground state for
$0.1<t_{3}/t<1$ is the same as for $t_{3}/t=1$.
For $t_{3}/t=0.1$, we can see the short $\log(4)$-plateau in the high
temperature region which stems from the degrees of freedom of the spin $S=1/2$ on the half-filled
3-site Hubbard chain and the spin $S=1/2$ on the site-3 regarded as
isolated from the Hubbard chain in this temperature region (a
separable state of the composite system of the 3-site Hubbard model and site-3).
But as the temperature slightly decreases from this region, we observe the
$\log(3)$-plateau due to the Lieb's ferrimagnetic state, which is finally
screened
in two steps by the Kondo effect in the same manner as the previous results.
We confirmed that 
this scenario for entanglement
($\log(4)\rightarrow \log(3)\rightarrow \log(2)\rightarrow 0$)
from the high temperature region toward the ground state 
for $t_{3}/t=0.1$  is the same as for $0.01<t_{3}/t\le 0.1$.
Alternatively, for $t_{3}/t=0.01$, the value of the impurity entropy
directly decreases from the $\log(4)$-plateau to the $\log(2)$-plateau
without via the $\log(3)$-plateau
as the temperature decreases from the high temperature region.
This is because the spin $S=1/2$ on the half-filled 3-site Hubbard chain
is at first screened by the Kondo effect (an entanglement state of the
composite system of the 3-site Hubbard chain and two reservoirs) and the spin $S=1/2$ on 
site-3 remains and generates the $\log(2)$-plateau. 
Therefore, it is suggested that we have $g_{s}\simeq 2e^{2}/h, \ 
g_{p}\simeq 2e^{2}/h$ in the
temperature region with the $\log(2)$-plateau.
However, as the temperature decreases further, the spin $S=1/2$ on the site-3 is
screened by electrons in the rest of the system coming through the renormalized hopping matrix element
$\tilde{t}_{3}$. The impurity entropy finally converges to zero, reaching the non-magnetic ground state.
This scenario for entanglement 
($\log(4)\rightarrow \log(2)\rightarrow 0$)
from the high temperature region toward the ground state
is expected for $0<t_{3}/t\le 0.01$.

The two-step Kondo screening for $0<t_{3}/t<0.01$ relates to that seen in
side-coupled double quantum dot systems 
\cite{PhysRevB.71.075305,PhysRevB.72.165309,PhysRevB.77.035120,PhysRevB.82.161411,PhysRevB.84.035119,PhysRevB.102.085418}
in the following aspects.
The side-coupled double quantum dot system consists of quantum dot A connected to two reservoirs and quantum dot B connecting or interacting only with quantum dot A.
The 1/2-spin on dot A and the 1/2-spin on dot B behave independently in the high temperature region.
As the temperature is decreased, screening of the 1/2-spin on dot A is first observed, followed by a subsequent screening of the 1/2-spin on dot B.
Meanwhile, the two-step Kondo screening
for $t_{3}/t>0.01$ is intrinsically different from the Kondo screening mentioned above.
This is because Lieb's ferrimagnetic state, which is an entangled
spin-triplet state between site-3 and sites-1, 2, and 4, is quenched for $t_{3}/t>0.01$. 

From these facts, we find that the $t_{3}/t$-independent ground state
for $t_{3}/t>0$, resulting from 
the $t_{3}/t$-dependent scenarios for the screening of the degrees of
freedom in high temperature regions,  
causes the robustness of the behaviors of $g_{s}$ and $g_{p}$ at the half-filled state.

\section{Conclusion}\label{last}
In summary, we investigated the Kondo effect on the minimal Lieb's
 ferrimagnetism on the T-shaped lattice connected to electron reservoirs
 by using a reliable method.
We found that the Lieb's ferrimagnetic moment $S=1$ 
 is screened in two steps by the Kondo effect.
Here we estimate one of our Kondo temperatures in units of Kelvin using recent experimental
values. In our calculations, Kondo temperatures are scaled by $D$ and we set $t/D=0.1$. 
The value of $t$ can be controlled from the $\mu$eV to sub meV in
recent experiments\cite{Natn,hensgens2017quantum}.
This range corresponds to $D$ being on the order of meV$\sim$10 K. Therefore,
even in the first step of the Kondo screening, the Kondo temperature
estimated from Fig. 5 for $U/t=2$ is to the order of 1 $\mu$K.
Therefore, the Kondo temperatures
 are
very low in the present case because we chose a relatively small
$\Gamma/t$ and a larger $U/t$. 
However, the Kondo temperature
rises as $\Gamma/t$ increases and $U/t$
decreases, which would make
the value of the Kondo temperature 
an accessible value in experiments, 
as shown in Fig.\ref{entro} and Fig.\ref{Gentro}.
In spite of the existence of the Kondo screening, we found that 
the conductance $g_{s}$ 
 is
 strongly suppressed $g_{s}\simeq 0$ while the conductance $g_{p}$ 
 has the maximum value $g_{p}\simeq 4e^{2}/h$.  
For these behaviors, we proposed one possible reason which should be confirmed
by further calculations, where the spin correlations among the sites of the
T-shaped lattice connected to the reservoirs would be clarified.
%
We also discussed the robustness of these behaviors of the conductance against the
perturbation toward reducing the Lieb's ferrimagnetism.
This robustness is caused by the perturbation-strength-independent ground state resulting from 
three perturbation-strength-dependent scenarios for entanglements of the degrees of freedom in high temperature regions.
%
It would be interesting that experimental investigations of the above mentioned properties of the Kondo
effect on the minimal Lieb's ferrimagnetism will be carried out in the future.
\begin{acknowledgements}
The authors acknowledge the fruitful discussions with 
Dr.A.C.Hewson, M.Sc.N.Shimada and M.Sc.M.Watanabe.
One of us (M.T.) acknowledges the support by JSPS
KAKENHI Grant No.JP20J20229.
Numerical computation was partly carried out in Yukawa Institute
Computer Facility.
\end{acknowledgements}

\bibliography{
bibtex
}

\begin{thebibliography}{32}
\bibitem[{exp()}]{expnon}
\bibinfo{note}{Here we mention the meaning of minimal and nontrivial Lieb's
  ferrimagnetism in our context. The minimal bipartite lattice is of course the
  two-site chain lattice decomposed into $N_{A}=1$ and $N_{B}=1$. However, the
  ground state of the half-filled and repulsive Hubbard model on the two-site
  chain lattice is non-magnetic state $S_{G}=0$. The three-site chain lattice
  is a bipartite lattice decomposed into $N_{A}=2$ and $N_{B}=1$, and the
  half-filled and repulsive Hubbard model on the three-site chain lattice has
  the magnetic ground state with $S_{G}=1/2$. However, the ground state
  including three electrons occupying in each the different three sites, which
  is naively expected for the half-filled Hubbard model with (large) repulsive
  interactions, is always a magnetic state. In this sense, the part of the
  statement for the existence of the magnetism in Lieb's theorem for this
  system is trivial.}
\end{thebibliography}
%
%
\bibliographystyle{apsrev}


\end{document}